\documentclass[twocolumn,english,xlinenumbers,aps,twocolumn]{revtex4-1}
\usepackage[T1]{fontenc}
\usepackage[utf8]{inputenc}
\usepackage{color}
\usepackage{babel}
\usepackage{verbatim}
\usepackage{amsmath}
\usepackage{graphicx}
\usepackage[unicode=true,pdfusetitle,
 bookmarks=true,bookmarksnumbered=false,bookmarksopen=false,
 breaklinks=false,pdfborder={0 0 0},pdfborderstyle={},backref=false,colorlinks=true]
 {hyperref}
\hypersetup{
 citecolor=blue}
\usepackage{breakurl}



\begin{document}

\title{A Functional Mapping for Passively Mode-Locked Semiconductor Lasers}

\author{C. Schelte$^{1,2}$, J. Javaloyes$^{1}$, S. V. Gurevich$^{2,3}$ }

\affiliation{$^{1}$ Departament de Física, Universitat de les Illes Balears, C/ Valldemossa km 7.5, 07122 Mallorca, Spain}
\affiliation{$^{2}$ Institute for Theoretical Physics, University of Münster, Wilhelm-Klemm-Str. 9, 48149 Münster, Germany}
\affiliation{$^{3}$ Center for Nonlinear Science (CeNoS), University of Münster, Corrensstrasse 2, 48149 Münster, Germany}

\begin{abstract}
We present a novel approach for the analysis of passively mode-locked
semiconductor lasers that allows for efficient parameter sweeps and
time jitter analysis. It permits accessing the ultra-low repetition
rate regime where pulses become localized states. The analysis including
slow (e.g. thermal) processes or transverse dynamics becomes feasible. 
Our method bridges the divide between the phenomenological,
yet highly efficient, pulse iterative model that is the Haus master
equation, and the more involved first principle descriptions relying
on time delayed equations. Our iterative functional mapping exploits
the fundamental division of the mode-locking regime between fast
and slow stages and allows computing the dynamics only in the pulse
vicinity. Reductions of the simulation times and of the memory footprint up
to two orders of magnitudes are demonstrated. Finally, the mapping also 
provides a general framework for deducing the Haus master equation
from first principle models based upon delayed differential equations.
\end{abstract}

\maketitle

Generation of low repetition rate picosecond pulses is of paramount
importance for a number of applications \cite{KT-PR-06,lidar}. Passive
Mode-locking (PML) of semiconductor lasers is a most promising method,
although it still represents an experimental and a theoretical challenge,
see \cite{AJ-BOOK-17} for a recent review. The Haus master equation
\cite{haus00rev} is an efficient and widely used approach to study
PML. It consists in restricting the analysis of the field to a small
temporal interval around the pulse. Yet this method, when applied
to a particular design, provides only qualitative predictions due
to the many simplifying hypothesis involved. How to derive \emph{the}
Haus equation, for a specific laser design, is also an open question.
On the other hand, first principle modeling allows representing the
full dynamics of both unidirectional and bidirectional cavities as
either Delay Differential Equations (DDEs) \cite{VT-PRA-05} or Delay
Algebraic Equations (DAEs) \cite{MB-JQE-05}, respectively. Such models
have been applied successfully to the study of PML with saturable
absorber (SA), and extended to describe photonic crystals \cite{HBM-OE-10},
external optical feedback \cite{JNS-PRE-16}, optical injection
\cite{RHR-PRE-11,AHP-JOSAB-16}, frequency swept sources \cite{SKO-OE-13}, 
quantum dot lasers \cite{RBM-JQE-11}, nonlocal imaging conditions \cite{MJB-JSTQE-15}
and localized structures (LSs) \cite{MJB-PRL-14}.

The understanding of PML is limited by its strongly multiscale nature,
in which the field and the gain temporal features differ by several
orders of magnitudes. Even semiconductor mode-locked lasers, that
have fast gain recovery $\tau_{g}\sim1\,$ns, generate pulses of a
duration $\tau_{p}\sim1\,$ps, hence $\tau_{p}/\tau_{g}\ll1$. Finding
the optimal operating regimes and studying the pulse train's amplitude
and temporal jitter require simulations over hundreds of thousands
of round-trips. If, in addition to the longitudinal dynamics, transverse
diffraction of the beam or slow (e.g. thermal) processes are taken
into account, the problem becomes quickly intractable. External cavity
devices are of particular interest for low frequency PML. These configurations
rely on vertical external-cavity surface-emitting semiconductor lasers
(VECSELs) coupled to distant saturable absorber mirrors SESAMs \cite{hoogland00,haring01,haring02,tropper04}.
Recently, a regime of temporal localization allowing arbitrary low
repetition rates was demonstrated in such systems \cite{MJB-PRL-14,CJM-PRA-16,JCM-PRL-16}.
In this regime in which the cavity round-trip $\tau$ is much larger
than $\tau_{g}$, the PML pulses become individually addressable temporal
LSs, that could also evolve into three-dimensional light bullets \cite{J-PRL-16,GJ-PRA-17}
if broad-area VECSELs are considered. However, the analysis in the
regime $\tau\gg\tau_{g}\gg\tau_{p}$ is particularly tedious. All
these arguments call for the development of more efficient methods.

In this manuscript, we present an approach that combines the accuracy
of first principle time-delayed models with the computational efficiency
of the Haus master equation. Our method is based upon computing only
the so-called \textit{fast stage} in the vicinity of the pulse where
stimulated emission is dominant, while using the analytical solution
of the dynamics during the \textit{slow stage} in-between pulses.
In the latter, pumping and carrier recombination are the dominant
processes, and the dynamics essentially consists in the exponential
recovery of the gain and of the absorption. Further, one reconnects this
analytically found slow-stage solution to the next round-trip's fast
stage as a connecting boundary condition. This idea is general and
it is applicable to any model described by DDEs or DAEs, and generally
to all PML lasers in which fast and slow stages can be identified. 

We illustrate the idea of the functional mapping (FM) using the DDE
model of \cite{VT-PRA-05} that considers unidirectional propagation
in a ring laser. The equations for the field amplitude $A$, the gain
$G$ and the absorption $Q$ read
\begin{eqnarray}
\frac{\dot{A}}{\gamma} & = & -A+Y\left(t-\tau\right),\label{eq:VT1}\\
\dot{G} & = & \Gamma\left(G_{0}-G\right)-e^{-Q}\left(e^{G}-1\right)\left|A\right|^{2},\label{eq:VT2}\\
\dot{Q} & = & Q_{0}-Q-s\left(1-e^{-Q}\right)\left|A\right|^{2},\label{eq:VT3}
\end{eqnarray}
with 
\begin{eqnarray}
Y\left(t\right) & = & \sqrt{\kappa} \exp\left[\frac{1-i\alpha}{2}G\left(t\right)-\frac{1-i\beta}{2}Q\left(t\right)\right]A\left(t\right),\label{eq:VT4}
\end{eqnarray}
and $G_{0}$ the pumping strength, $\Gamma=\tau_{g}^{-1}$ the
gain recovery rate, $Q_{0}$ the value of the unsaturated losses which
determines the modulation depth of the SA and $s$ the ratio of the
saturation energy of the gain and of the SA sections. We define $\kappa$
as the intensity transmission of the output mirror, i.e., the fraction
of the power remaining in the cavity after each round-trip. In Eqs.~(\ref{eq:VT1}-\ref{eq:VT3})
time has been normalized to the SA recovery time that we assume to
be $\tau_{q}=20\,$ps. The linewidth enhancement factor of the gain
and absorber sections are noted $\alpha$ and $\beta$, respectively.
In addition, $\gamma$ is the bandwidth of the spectral filter whose
central optical frequency has been taken as the carrier frequency
for the field. This spectral filter may (coarsely) represent, e.g.,
the resonance of a VCSEL \cite{MJB-JSTQE-15}. If not otherwise stated
$\left(\kappa,\alpha,\beta,s\right)=\left(0.8,2,0.5,30\right)$, and
$Q_{0}=0.3$ which corresponds to modulation of the losses of $\sim26\,\%$.
We also set $\gamma=10$ and $\Gamma=0.04$, leading to a gain bandwidth
full width at half maximum of $160\,$GHz and $\tau_{g}=500\,$ps.

\begin{figure}
\centering{}\includegraphics[clip,width=1\columnwidth]{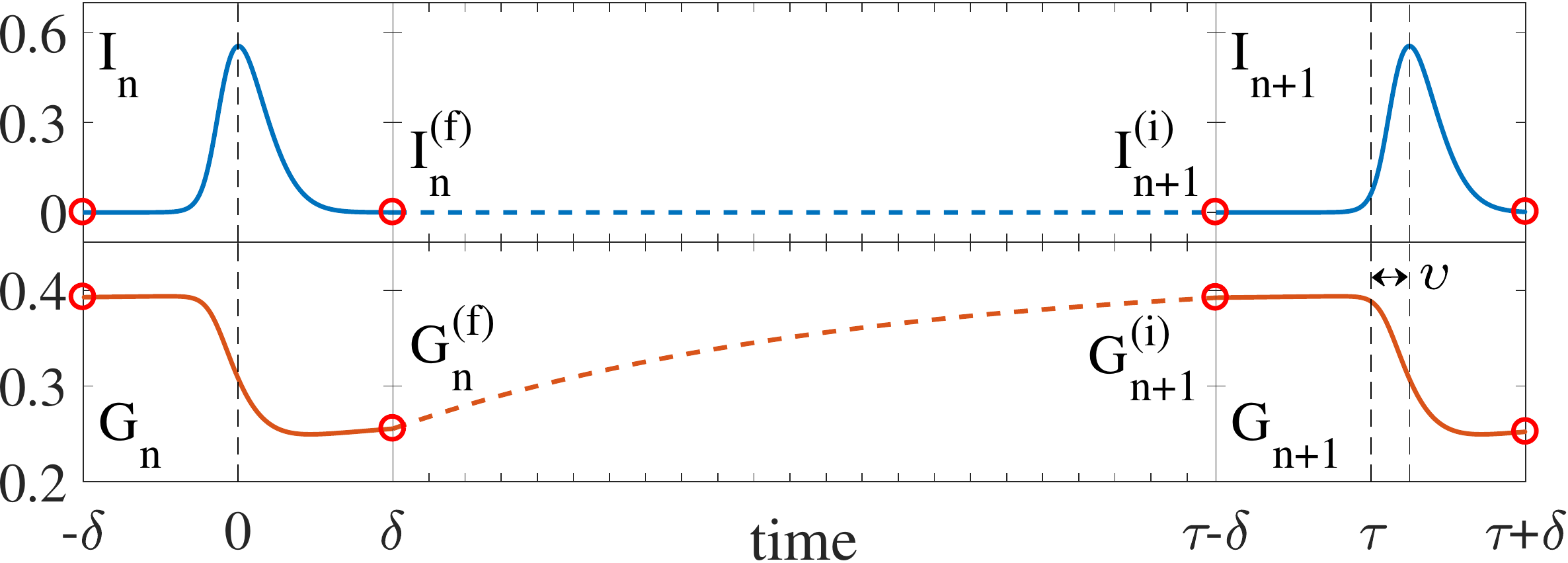}\caption{Temporal profile of the intensity $I_{n}=\left|A_{n}\right|^{2}$
and of the gain $G_{n}$ at the $n$-th and $n+1$-th round-trips.
After the emission of the pulse $I_{n}$ and the ensuing gain depletion,
the so-called fast stage (solid lines), the gain recovers until the
next round-trip while the field is vanishing (dashed lines). Knowing
the final value of $G_{n}^{\left(f\right)}$ in the interval $z\in\left[-\delta,\delta\right]$,
one can deduce the initial gain value at the next round-trip $G_{n+1}^{\left(i\right)}$.
The central panel (not up to scale) can be several orders of
magnitude larger than the outer panels.}
\label{fig:DIAG}
\end{figure}

We wrote Eq.~(\ref{eq:VT1}) in a form that makes apparent that the
forcing field $Y\left(t-\tau\right)$ defined in Eq.~(\ref{eq:VT4})
is a nonlinear function that is known over a whole interval of duration
$\tau$. Since $G$ and $Q$ are functions of $A$, $Y$ involves
\emph{only} the past values of the field, i.e., $Y\left(t-\tau\right)=g\left[A\left(t-\tau\right)\right]$.
Knowing the forcing term $Y$, Eq.~(\ref{eq:VT1}) can be solved
for $A$ over an interval of duration $\tau$. Integrating Eqs.~(\ref{eq:VT1}-\ref{eq:VT4}),
not over the whole round-trips, but only in a selected time interval
in the pulse vicinity, is at the core of our method. We define the
field and carrier profiles at the $n$-th round-trip as $A_{n}\left(z\right)$
and $\left(G_{n},Q_{n}\right)\left(z\right)$. For clarity, we set
in Fig.~\ref{fig:DIAG} the pulse at the origin of time at the $n$-th
round-trip. Next, we define a small interval of duration $2\delta$
and a local time $z\in\left[-\delta,\delta\right]$. Finally, we impose
a condition on the waveform $A_{n}$: it is a pulse of duration $\tau_{p}$
asymptotic to $A=0$ if $\delta\gg\tau_{p}$. Under these approximations,
one can solve Eq.~(\ref{eq:VT1}) using standard integration techniques,
e.g., Runge-Kutta method, at the next round-trip, using the following
sequence $\left(A_{n},G_{n},Q_{n}\right)\rightarrow Y_{n}\rightarrow\left(A_{n+1},G_{n+1},Q_{n+1}\right)$.
Doing so corresponds to writing a functional mapping $A_{n+1}=F\left(A_{n}\right)$.
The remainder of the dynamics during the round-trip of duration $r=t-2\delta$,
see central panel in Fig.~\ref{fig:DIAG}, in which the field is
vanishing can be found by solving Eqs.~(\ref{eq:VT2},\ref{eq:VT3})
analytically in the absence of stimulated emission, the so-called
slow stage of PML. As such, $G_{n+1}^{(i)}=G_{n}^{(f)}\chi+G_{0}\left(1-\chi\right)$
with $\chi=\exp\left(-\Gamma r\right)$ and similarly for $Q_{n+1}^{(i)}$.
Solving analytically the slow stage allows to fully cancel the stiffness
inherent to the multiscale nature of PML which is exceptionally useful
in the long delay limit $\tau\gg\tau_{g}$. The speedup of our method
is equal to the ratio of the actual integration domain $2\delta$
and of the full round-trip $\tau$, i.e., $m=\tau/\left(2\delta\right)$.
Taking a domain of duration $2\delta=5\tau_{p}$, a pulse-with of
$\tau_{p}=1\,$ps at a repetition rate of $\tau^{-1}=1\:$GHz, yields
a speedup $m=200$, i.e., a 24 hour simulation with, e.g., slow thermal
effects or transverse diffraction could potentially be achieved in
a few ($\sim7$) minutes. 
We stress that our method can potentially be extended to the case of a non zero background field.
Such a situation could occur, e.g., in presence of monochromatic optical injection and in this case 
the PML pulses would exist on a non zero background field as in the Lugatio-Lefever model \cite{LL-PRL-87}.
However, this background homogeneous state upon which the pulses connect asymptotically 
would need to be unconditionally stable.

Time-delayed systems are essentially convective \cite{GP-PRL-96}.
As such, the period of the solution is always slightly different from
the value of the time delay and the pulse $A_{n+1}$ at the next round-trip
will be shifted, see Fig.~\ref{fig:DIAG} right panels. In the case
of PML, this drift admits an intuitive interpretation: If the pulse
at the $n$-th round-trip $A_{n}$ is centered, the next iterate of
the pulse $A_{n+1}$ will slightly by shifted to the right, of an
amount $\upsilon=\gamma^{-1}$, as a consequence of the inertia of
the resonant filter. An effect modeled by the parameter $\gamma$
that represents for the case of a VCSEL, the time the photons remain
trapped in the cavity. The pulse can also deviate from the center
of the interval due to stochastic fluctuations. Hence, one needs to
recenter the pulse at each round-trip leading directly to the value
of the pulse jitter.

Our method is, in addition, a rigorously way to deduce the Haus master
equation in a general setting. For the case of Eqs.~(\ref{eq:VT1}-\ref{eq:VT4}),
one can use the Fourier transform $\mathcal{F}$ yielding an explicit,
expression of the mapping operator
\begin{eqnarray}
A_{n+1} & = & \mathcal{F}^{-1}\left[\mathcal{L}\left(\omega\right)\mathcal{F}\left[g\left(A_{n}\right)\right]\right],\label{eq:Sol_om}
\end{eqnarray}
where we defined the Lorentzian kernel $\mathcal{L}\left(\omega\right)=\left(1+i\omega/\gamma\right)^{-1}$
. The fact that equations of the type as Eq.~(\ref{eq:VT1}) could
be solved by a Fourier method was already pointed out in \cite{GP-PRL-96}
for the case of a single variable. The nominal drift can easily be
accounted for setting $\mathcal{L}\left(\omega\right)\rightarrow\mathcal{L}\left(\omega\right)\exp\left(i\omega\upsilon\right)$
with $\upsilon=\gamma^{-1}$. Introducing a slow time scale for the
evolution of the field after each round-trip as $\partial A/\partial\xi=A_{n+1}-A_{n}$,
we find
\begin{eqnarray}
\frac{\partial A}{\partial\xi} & \equiv & \mathcal{F}^{-1}\left[\mathcal{L}\left(\omega\right)\mathcal{F}\left[g\left(A_{n}\right)\right]\right]-A_{n}.\label{eq:Real_Haus}
\end{eqnarray}
By inspecting Eq.~(\ref{eq:Real_Haus}), one can notice how obtaining
the Haus equation necessitates a wealth of approximations. One needs
to assume the uniform field limit (UFL), i.e. small gain, absorption,
and losses, and phase change. In addition, we expand in Taylor
series $\mathcal{L}\left(\omega\right)$ and truncate to second order
in $\omega$. Finally, we find setting $i\omega\rightarrow\partial/\partial z$
\begin{eqnarray}
\frac{\partial A}{\partial\xi} \hspace{-0.3cm}& = &\hspace{-0.3cm} \left\{ \sqrt{\kappa}\left(1+\frac{1-i\alpha}{2}G-\frac{1-i\beta}{2}Q\right)-1+\frac{1}{2\gamma^{2}}\frac{\partial^{2}}{\partial z^{2}}\right\} A.\label{eq:Shit_Haus}
\end{eqnarray}

It was under these approximations that the Haus equation was
derived from Eqs.~(\ref{eq:VT1}-\ref{eq:VT3}) using the multiple
time scales formalism in \cite{KNE-PD-06,CJM-PRA-16}. We also note
that even in the UFL, the Haus equation remains an approximation. 
A continuous dynamical system can not emulate correctly
a discrete mapping. In addition, writing a parabolic partial
differential equation remains an approximation of the time delayed
operator $\mathcal{L}\left(\omega\right)$ described by a Lorentzian, 
only valid for narrow bandwidth fields. 

\begin{figure}
\centering{}\includegraphics[clip,width=1\columnwidth]{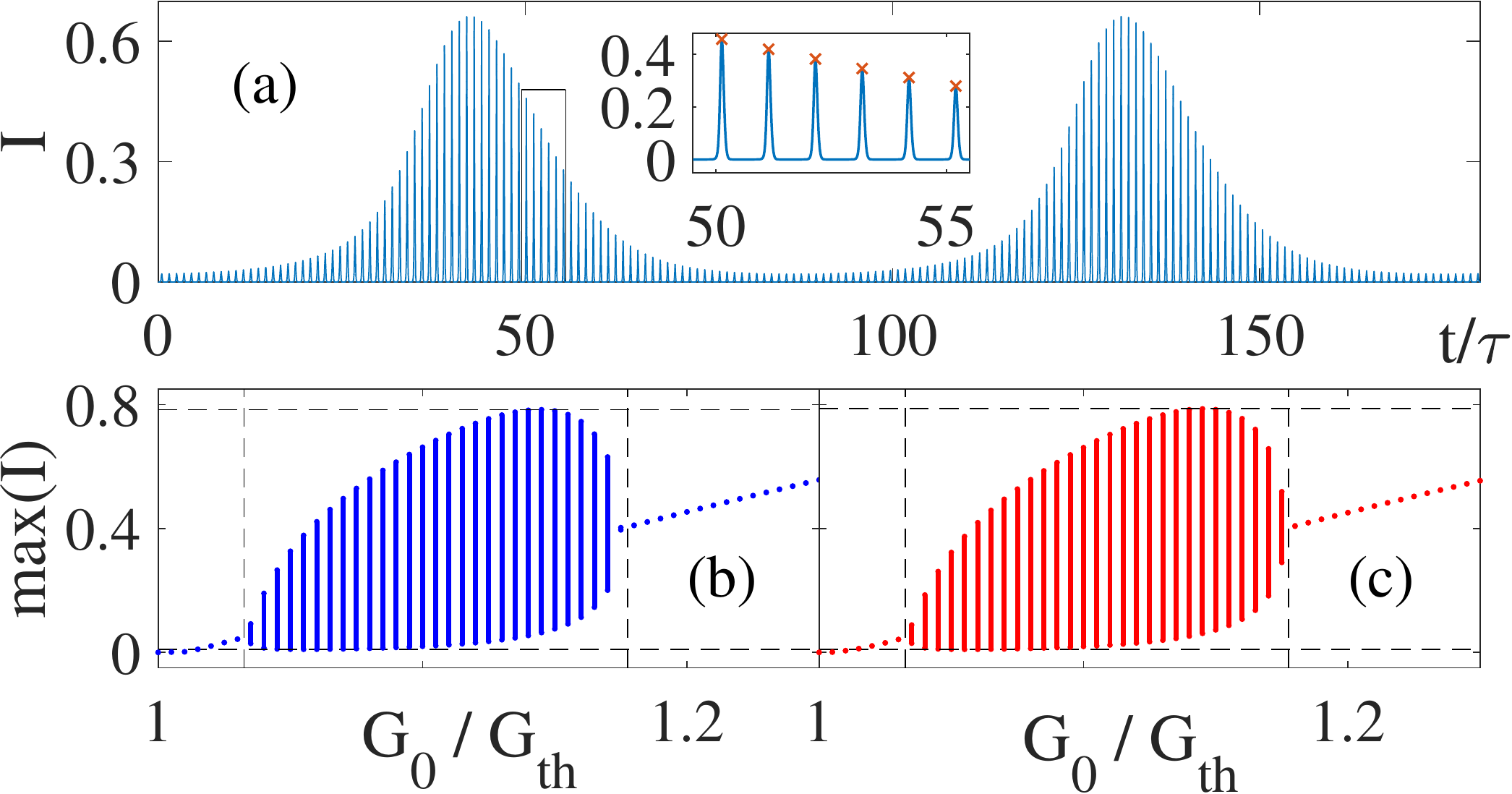}\caption{(a) Temporal trace of the self-pulsing regime found close to the lasing
threshold $G_{th}$ , obtained for $\tau=2$ and $Q_{0}=0.6$ . The
bifurcation diagrams recording the pulses intensity maxima are represented
in (b) for the DDE model (\ref{eq:VT1}-\ref{eq:VT4}) and (c) for
the FM (\ref{eq:Sol_om}). Residual differences can be found.}
\label{fig:SP}
\end{figure}

The FM can be used to study the dynamics of the pulse, like the leading and 
trailing edge instabilities, and in general the unstable regimes 
where the pulse breathes in height and width. At variance with the Haus 
equation, we will show in Fig.~\ref{fig:SP} that the gain induced instabilities, 
such as self-pulsation, are also conserved due to the proper consideration of 
the gain recovery dynamics.
However, we note that the FM suffers from the same limitation than all pulse
iterative models, i.e., the modal structure of the laser is lost and so are 
the transitions toward harmonic PML, so that in the FM, it is essential to 
assume the background stability criterion. However, we note that this potential 
weakness can be mitigated since the Harmonic transition is relatively easy to predict 
by monitoring the maximal value of the net gain during its recovery. 
As long as it remains negative, the spontaneous transitions toward Harmonic PML is inhibited.
\begin{figure}
\centering{}\includegraphics[clip,width=1\columnwidth]{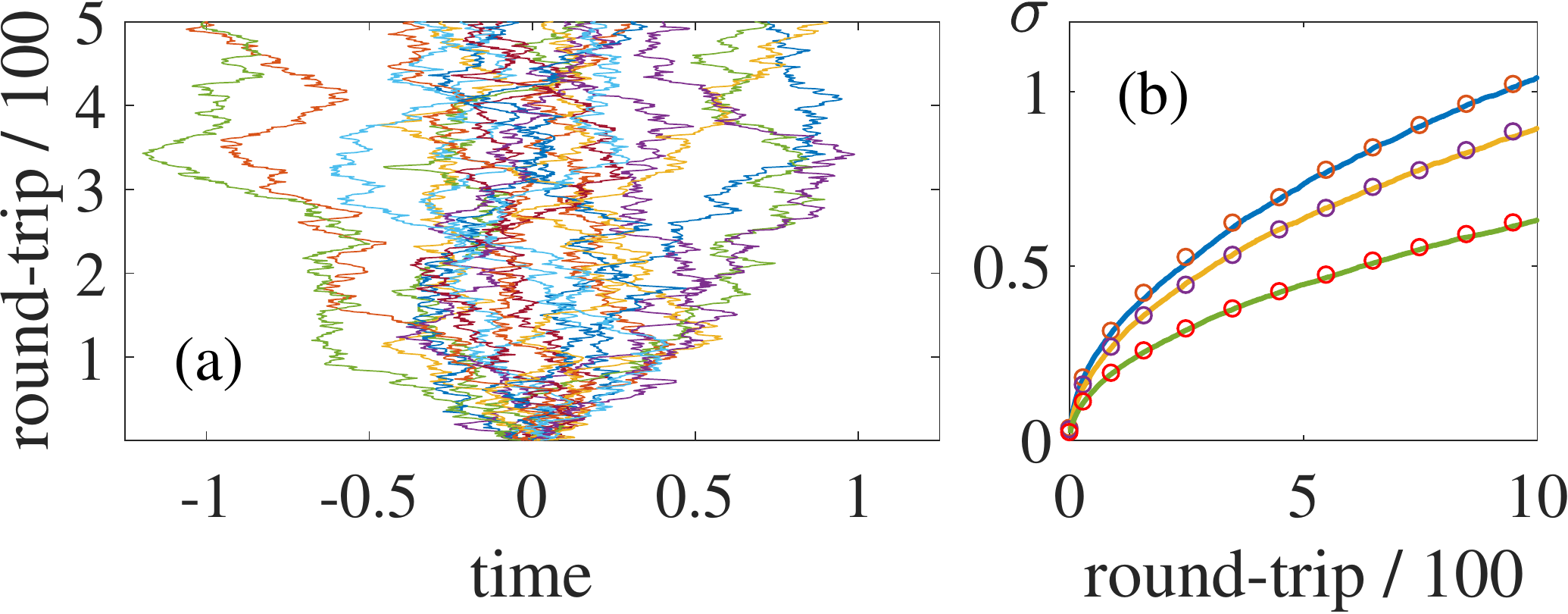}\caption{Temporal jitter of the pulse in the fundamental PML regime. (a) Twenty different realizations leading to uncorrelated random walks. (b) Variance
of the random walk for different parameters that correspond from top
to bottom to the short i), medium ii) and long iii) cavity regimes,
see text for details. The lines and the circles correspond to the DDE 
model~(\ref{eq:VT1}-\ref{eq:VT4}) and the FM~(\ref{eq:Sol_om}), respectively.}
\label{fig:jitter}
\end{figure}

While the stiffness of PML could, in principle, be alleviated by the use of 
adaptive time-step method, these algorithms do not allow for stochastic analysis and
to include transverse effects such as diffraction, which can easily
be studied with the FM. 
We show in Fig.~\ref{fig:jitter} the results of a time jitter analysis,
either integrating Eqs.~(\ref{eq:VT1}-\ref{eq:VT4}) or the FM given 
by Eq.~(\ref{eq:Sol_om}). For the sake of simplicity,
we added white delta correlated noise of variance $\sigma_{e}=10^{-2}$
in the field equation only, but similar fluctuations could be introduced
in the carrier equations to model current fluctuations. We performed
statistics in different regimes, the short i) $\left(G_{0},\tau\right)=\left(0.6,5\right)$,
average ii) $\left(G_{0},\tau\right)=\left(0.5,10\right)$, and long
iii) $\left(G_{0},\tau\right)=\left(0.4,100\right)$ cavities. The
domain length hosting the pulse in the FM is $2\delta=3.$ Statistics
were performed averaging the pulse jitter over $N_r=10^{4}\,$realizations
of $10^{3}\,$round-trips. A few trajectories are depicted in Fig.~\ref{fig:jitter}
(a) to exemplifies the random walk that the pulse performs from one
round-trip toward the next while Fig.~\ref{fig:jitter} (b) depicts
the statistical variance $\sigma$ of the pulse distribution, that
grows as $\sigma=\sqrt{2Dt}$ as predicted by the theory of Brownian motion.
We notice that the FM yields identical results for the diffusion coefficient
than the DDE model~(\ref{eq:VT1}-\ref{eq:VT4}). However, such jitter results
can be obtained in a few minutes using the FM, and the time necessary
does not increase with the value of $\tau$. The agreement between
the two approaches stem from the fact that the neutral mode that
corresponds to the translation of the pulse in time is correctly preserved by the FM. 
We note that the value of $\sigma_{e}$ used leads to very large jitter, demonstrating 
that both methods are in agreement, even for large pulse to pulse deviations 
($\sim 10\,\%$ in peak intensity). Large $\sigma_{e}$ allows to mitigate the 
finite size fluctuations proportional to $1/\sqrt{N_r}$  upon averaging over realizations. 
Performing a similar comparison for lower values of $\sigma_{e}$ would require 
prohibitively longer integration of Eqs.~(\ref{eq:VT1}-\ref{eq:VT4}). 
Here, the semi-analytical method of \cite{JPR-PRA-15} should be used instead to compare with the FM.

In the long cavity limit that allows for temporal pulses to become
temporal LSs \cite{MJB-PRL-14,JCM-PRL-16}, all the bifurcation diagrams
we performed as a function of all possible parameters led to a perfect
superposition between the DDE model and the FM, the later leading
to greatly reduced integration time. While in the short cavity regimes,
i.e., $10\,$GHz and beyond, the use of the FM leads to less marked
improvements, we demonstrate in Fig.~\ref{fig:SP} that even the
self-pulsation region found in the vicinity of the threshold is preserved.
The self-pulsation (SP) regime is found in high frequency semiconductor
mode-locked lasers and it consists in relaxation oscillations between
the pulse energy and the population inversion. A proper consideration
of the carrier dynamics from one round-trip toward the next is essential
to reproduce the dynamics of SP, that is not found in the Haus equation.
We show in Fig.~\ref{fig:SP} (a) a temporal trace of the SP regime
found close to the lasing threshold $G_{th}=Q_0-\log \kappa$, whereas the bifurcation
diagrams showing the pulses intensity maxima are represented in panels
(b) and (c) for the DDE (Eqs.~\ref{eq:VT1}-\ref{eq:VT4}) and the FM (Eq.~\ref{eq:Sol_om}),
respectively. One can clearly see that the onset and disappearance of SP
is well preserved by the FM, since the correlations between the gain
and the field intensity from one round-trip to the next are properly
accounted for. Here, $\tau=2$ and $Q_{0}=0.6$ which corresponds
to a $25\,$GHz repetition rate and a $45\,\%$ modulation of the
losses. The domain size in the FM is $2\delta=1.5.$ We note that
for high frequency PML other modeling approaches such as Traveling
Wave Models \cite{JB-JQE-10,JB-PRA-10,Freetwm} are better indicated.

We conclude our analysis by showing how the FM can be used for the
simulation of broad area MIXSEL system described by the DAE model
of \cite{MB-JQE-05,MJB-JSTQE-15}. The model for the intra-cavity
field $E$, gain $N_{1}$ and absorption $N_{2}$ reads
\begin{eqnarray}
\dot{E} & = & \left[\left(1-i\alpha_{1}\right)N_{1}+\left(1-i\alpha_{2}\right)N_{2}-1+i\Delta_{\perp}\right]E+hY,\label{eq:MIXSEL1}\\
\dot{N}_{1} & = & \gamma_{1}\left(J_{1}-N_{1}\right)-N_{1}\left|E\right|^{2}\,,\label{eq:MIXSEL2}\\
\dot{N}_{2} & = & \gamma_{2}\left(J_{2}-N_{2}\right)-sN_{2}\left|E\right|^{2}\,.\label{eq:MIXSEL3}
\end{eqnarray}
while the relation linking the intra-cavity and the external cavity $Y$ fields 
is, after proper normalization that considers the cavity enhancement factor,
$Y=\eta\left[E\left(t-\tau\right)-Y\left(t-\tau\right)\right]$ with
$\eta$ the external mirror reflectivity. The coupling of $Y$ into
the MIXSEL cavity is given by the parameter $h$, see \cite{MJB-JSTQE-15}
for more details. We operate in the long cavity limit, such that $\chi=0$
and the value of $\tau$ is irrelevant. For the sake of simplicity,
we only consider diffraction in a single transverse dimension, making the
problem two-dimensional and allowing for easier multi-parameter bifurcation
analysis. We concentrate on the spatio-temporal localization regime.
Here, the PML pulses become temporal dissipative solitons that 
can be addressed individually, but at the same time, they also acquire 
a well defined transverse size. This regime where the field coalesces into 
a spatio-temporal soliton is called a Light Bullet (LB) \cite{J-PRL-16,GJ-PRA-17}. 

\begin{figure}
\centering{}\includegraphics[bb=0bp 15bp 509bp 200bp,clip,width=1\columnwidth]{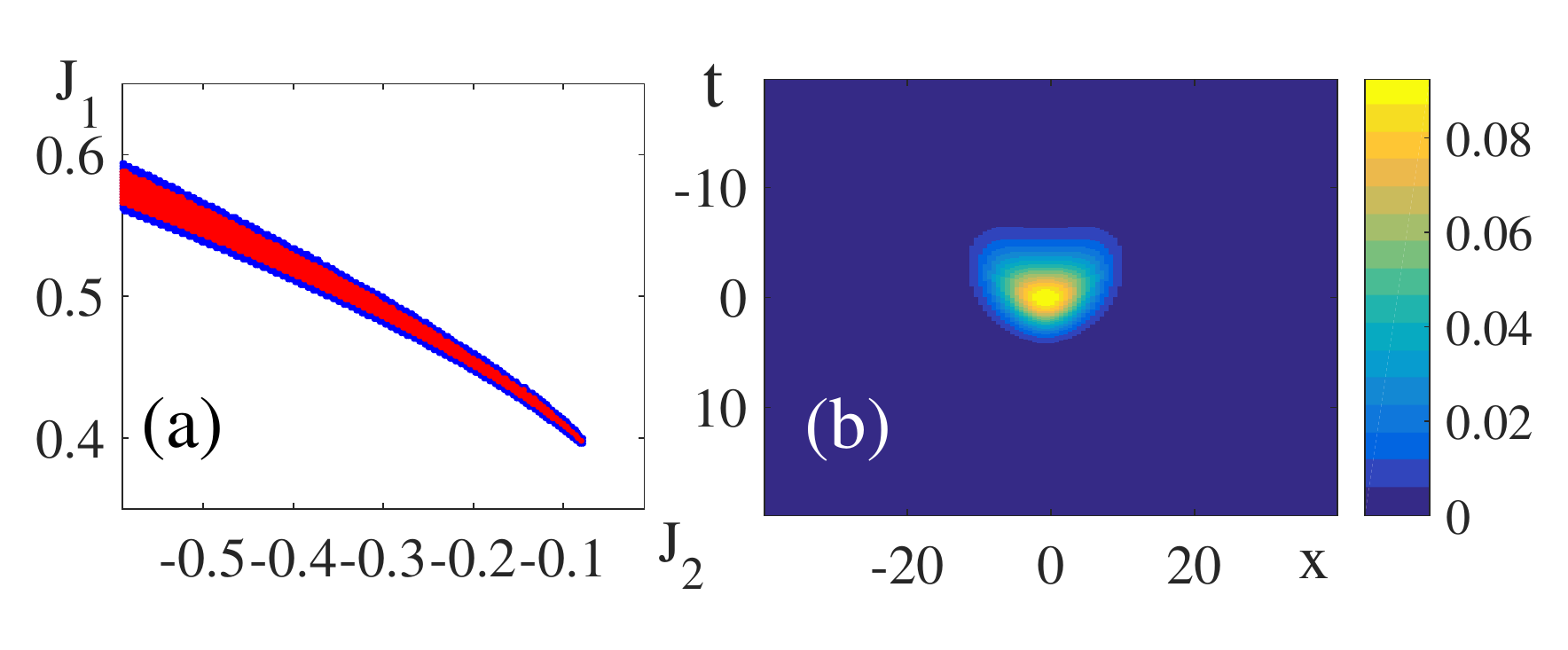}\caption{(a) Two-dimensional bifurcation diagram showing the region of stable
existence of the LBs of the DAE model (\ref{eq:MIXSEL1}-\ref{eq:MIXSEL3})
as a function of the reverse bias in the gain and absorber sections
$J_{1}$ and $J_{2}$. (b) Spatio-temporal profile of the field found
with the FM with $\left(J_{1},J_{2}\right)=\left(0.498,-0.336\right)$.
Other parameters are $\left(\alpha_{1},\alpha_{2},h,\gamma_{1},\gamma_{2},s,\eta\right)=\left(0,0,2,0.003,0.1,30,0.5\right)$.}
\label{fig:LB}
\end{figure}

Figure~\ref{fig:LB} (a) shows a two-dimensional bifurcation diagram for the case $\alpha_{j}=0$
as a function of the bias in the gain and absorber sections $J_{1}$ and $J_{2}$. 
Finding the region of stability in
Fig.~\ref{fig:LB} (a) required 24 hours on a standard PC using the
FM, instead of several months integrating the full DAE system. Figure~\ref{fig:LB}
(b) depicts the spatio-temporal LB profile obtained with the FM method. 
Figure~\ref{fig:LB} (b) corresponds to a snapshot of the intracavity field with $t$ 
the temporal (cavity) axis and $x$ the transverse dimension.

In conclusion, we presented a modern approach for the analysis of
PML in semiconductor lasers based upon a functional mapping and demonstrated
its usefulness for parameter sweeps, time jitter analysis and spatio-temporal
dynamics. In particular, the ultra-low repetition rate regime where
the pulses become temporal LSs can be accessed easily, even in the
presence of transverse effects. Our method also provides a general
framework for deducing the Haus master equation from all models based
upon delayed differential equations. While the Haus equation can be
recovered in the uniform field limit, the FM possesses a stronger
regime of applicability, in particular in the regime of strong gain
and absorption, and we anticipate that significant deviations between
DDE or DAE models and the Haus equations can be obtained, which will
be the subject of further studies. Finally, we showed that the strong
reduction in the degrees of freedom is also useful for studying transverse
beam dynamics and more generally transverse patterns in PML. While
the dynamics in presence of transverse and/or slow effects such as
diffraction or thermal lensing is a topic of further studies, we gave
an example how a multi-dimensional bifurcation diagram for Light Bullets
can be obtained by means of the functional mapping method.

\section*{Funding information} 
MINECO Project COMBINA (TEC2015-65212-C3-3-P). 

\section*{Acknowledgments}
We acknowledge useful discussions with A. Vladimirov and A. Pimenov.

\end{document}